\documentclass [12pt]{article}
\usepackage{graphicx}
\usepackage{setspace}
\usepackage{amsmath}
\usepackage{xcolor}
\usepackage{float}
\usepackage{enumitem}
\usepackage{subcaption}
\title{Analysis of the Impact of Unforced Errors in Tennis}
\author{Hashan Peiris, Nirodha Epasinghege Dona and Tim B.\ Swartz
\thanks{
H.\ Peiris is a PhD candidate and T.\ Swartz is Professor,
Department of Statistics and Actuarial
Science, Simon Fraser University, 8888 University Drive, Burnaby
BC, Canada V5A1S6.
N.\ Epasinghege Dona is Senior Data Analyst, BC Health. 
Swartz has been partially supported by the
Natural Sciences and Engineering Research Council of Canada.
The work has been carried out with support from the CANSSI (Canadian Statistical
Sciences Institute) Collaborative Research Team (CRT) in Sports Analytics.
}}

\date{ }
\begin{document}
\maketitle
	
\begin{abstract}
\noindent
This paper considers the impact of unforced errors in sport. Although the 
proposed methods are applicable to various sports, we 
demonstrate the approach in the context of professional tennis.
The value of the approach is that we can provide estimates of
the points lost, games lost, sets lost
and matches lost due to unforced errors.
The methods are based on a bootstrapping procedure which also yields
standard errors for the estimates.
The approach is valuable in terms of player evaluation, and
can also be used for training purposes where it is
possible to assess the quantification of improvement
based on fewer unforced errors.
\end{abstract}
	
\vspace{3mm}

\newpage
\section{INTRODUCTION}

The term ``unforced error'' is often mentioned in sport. It is mistake made
by a player or by a team that is viewed as unnecessary and caused entirely by
the player or the team.
For example, in American football, a coaching staff has complete control
over the assignment of players on the field, and therefore, the penalty
``too many men on the field'' would be regarded as an unforced error.
Since unforced errors are controllable, they are regrettable, and 
are often discussed and criticized.

Despite the widespread acknowledgment that unforced errors ought to
be minimized, the analysis of unforced errors is a topic that has 
received minimal attention in the sports analytics literature.
In badminton, Yadav and Shukla (2011) considered 30 singles matches 
in university matches, and observed that
match winners had significantly fewer unforced errors than losers.
Again in badminton, Barreira and Chiminazzo (2020) considered 57 men's singles
matches from the 2016 Olympic Games, and observed that
match winners had significantly fewer unforced errors than losers.
In tennis, Brody (2006) outlines how players can reduce unforced errors
by adhering to various principles of physics.
Also in tennis, Attray and Attray (2021) provide a theoretical framework
for studying unforced errors. They suggest that the minimum number
of additional points needed to win a game given an unforced error
is a measure of the impact of an unforced error. Given varying match circumstances (i.e.,
the score),
unforced errors have impacts of varying magnitude.

We suggest that there are two reasons for the paucity of 
sophisticated studies involving unforced errors in matches:
\begin{enumerate}
\item The determination of whether a mistake was an unforced error or
a forced error is often subjective, and typical datasets do not provide
the distinction.
\item It is not always obvious how to evaluate the impact of an unforced error.
Had the unforced error not occurred, what follows in the match is a
counterfactual. That is, without the unforced error, it is unclear how the
remainder of the match would have unfolded. 
\end{enumerate}

In this paper, we consider the impact of unforced errors in tennis. 
Fortunately, we have access to a dataset that categorizes tennis errors
as either forced or unforced. 
The categorization is carried out by trained volunteers who watch video
recordings of matches. The training aspect is important as 
it reduces subjectivity and improves the reliability of the datasets. 
We overcome the counterfactual problem
by simulating the remainder of a match following an unforced error using
a bootstrapping procedure. We utilize the probabilities of what may have
happened had the unforced error not occurred.
By repeatedly simulating the remainder of the match had the unforced error
not occurred, we obtain the expected impact of the unforced error. 
Further, we can investigate ``what-if'' scenarios. For example, we might
pose the question, how would a player have fared in a match had the
player reduced unforced errors by 10\%? Answers to these types of questions are
valuable in terms of player evaluation and player development.

Although unforced errors do not appear to have been analyzed in tennis,
tennis has seen an upsurge of work in analytics.
One of the first serious statistical contributions to tennis analytics was the work by
Klassen and Magnus (2001) which concerned an investigation of the iid assumption that
points are independent and identically distributed. They concluded that there is
a positive correlation between successive outcomes and that servers are less likely to
win a point in important situations.

Baker and McHale (2014, 2017) used the classical Bradley-Terry model to compare the
performance of tennis players. McHale and Morton (2011) employed the same model for
the purpose of forecasting tennis matches.
Epasinghege Dona, Gill and Swartz (2024) 
investigate player characteristics in relation to to the length of the rally.
A comprehensive text on tennis analytics for Wimbledon is given by
Klaasen and Magnus (2014).

Sports analytics has been assisted by the availability of tracking data.
With tracking data, player and ball locations are recorded with high frequency
(i.e., spatio-temporal data), and these detailed datasets
have contributed to explorations of many sporting problems that were previously unimaginable.
Gudmundsson and Horton (2017) provide a review on spatio-temporal
analyses used in invasion sports where
player tracking data are available.
In tennis, there are a growing number of papers that provide statistical analyses
that rely on tracking data. 
Tea and Swartz (2023) investigate serving
tendencies which may allow a player to anticipate the nature of the opponent's serve.  
The approach relies on hierarchical models in a Bayesian framework.
Kovalchik and Albert (2022) also used a Bayesian framework where
they investigate serve returns by introducing a semiparametric mixture model.
Other papers 
that use tracking data but
focus exclusively on the serve include
Mecheri et al.\ (2016) and Wei et al.\ (2015).
The text by Albert et al.\ (2017) provides a flavour for sports statistics across major sports.

In Section 2, we describe the data and associated issues.
In Section 3, we describe the proposed bootstrapping methods that permit the
quantification of the impact of unforced errors. 
Specifically, we provide estimates of
the expected points lost, games lost, sets lost
and matches lost due to unforced errors.
In Section 4, we explore some descriptive statistics
and we illustrate the proposed methods in the context of a match between
tennis professionals.
We conclude with a short discussion in Section 5.
	
\section{TENNIS DATA}

The data utilized in this paper are taken from the 
nearly 1.5 million 
serves in the men's (ATP) and 
women's (WTA) professional tennis matches.
The data are available from the Match Charting Project 
(https://github.com/JeffSackmann/tennis\_MatchChartingProject)
maintained by Jeff Sackman. The data involve matches from 1970 through 2022
and contain shot-by-shot outcomes involving 710 distinct men and 489 distinct women.
This public dataset provides information on
shot type, shot direction, 
depth of returns, types of errors, and more. The data were collected by trained volunteers 
after watching the video recordings of matches. To the best of our knowledge, there is no other source
of publicly-available data of this type. At present, these data appear to be underutilized
with respect to statistical analyses.

The data cover matches from all the major Grand Slam events, the Davis Cup and many minor tournaments. 
The best players of this time period are included
in the dataset. Since the charting of matches was at the discretion of volunteers, data are 
skewed in favour of later years. Also, better players have a higher representation since they 
frequently reached the latter rounds of tournaments. 

\subsection{Data Management Issues}

When analyzing data, it is obviously important that the data are accurate.
Therefore, we carried out various procedures to check data accuracy.

In the MatchChartingProject dataset, rows correspond
to points awarded. Therefore,
we augmented the dataset to include all serves. For example,
whenever a second serve occurred, this implied that there was a first
serve that resulted in a fault, for which no point was awarded.

We observed minimal numbers of obvious data entry errors.
One example concerns the rallyCount variable where
there were five non-numeric characters in
the ATP data. We simply removed the corresponding rows when this occurred.

\section{METHODS}

Consider a match or a series of matches between player A and player B. We
wish to consider the impact of the unforced errors committed by player A in these matches.
Note that we only define unforced errors as those that lead directly to a point obtained
by player B; therefore first serve faults by player A are not considered unforced errors.

We categorize the serves into four datasets according to:
\begin{enumerate}
\item player A serving to player B on first serve
\item player A serving to player B on second serve
\item player B serving to player A on first serve
\item player B serving to player A on second serve
\end{enumerate}

We first consider the scenario where player A commits unforced errors
at the same rate that occurred in the datasets.
We do not provide the details of scoring in tennis but note that players accumulate
points which lead to games won which lead to sets won which lead to matches won.
The bootstrapping algorithm for this scenario is easily described
and begins with one of the players initiating service in Game 1 of Set 1.
As the simulation of a match progresses, the score of the match is updated
and the server is determined according to the rules of tennis.
For each serve, the simulation of a serve result proceeds as follows:
\begin{itemize}
\item determine the relevant dataset based on the score of the match
\item randomly sample one serve from the dataset
\begin{itemize}
\item record the outcome  
based on the sampled serve
\end{itemize}
\end{itemize}

The algorithm is simple and allows us to simulate a hypothetical match
where player A has made unforced errors at the same rate as in the datasets. 
We can then repeat the
algorithm over many matches to obtain the expected points won by
player A, the expected games won by player A, the expected sets won by player A and 
the expected matches won by player A.

We then consider a second and unrealistic scenario
where player A does not make unforced errors.
This permits the calculation of the same expected values as before.
Therefore, the differences in the expected values
under the two scenarios 
provides a quantification of the impact of unforced errors. 
The second scenario of no unforced errors provides
us with a baseline for the maximum amount of improvement
that a player can make.

The simulation algorithm in the second scenario is slightly more complicated
since it is unknown which player would have won the point had an unforced
error not occurred.
For each serve in this scenario, the simulation of a serve result proceeds as follows:
\begin{itemize}
\item determine the relevant dataset based on the score of the match
\item randomly sample one serve from the dataset 
\begin{itemize}
\item if player A does not commit an unforced error, record the outcome
based on the sampled serve
\item if player A does commit an unforced error, record the outcome
as generated from the probabilities in Table 1
\end{itemize}
\end{itemize}

Table 1 provides player point probabilities derived from the work of
Epasinghege Dona, Gill and Swartz (2024). 
The probabilities correspond to A winning the point had
A not made an unforced error on touch $t$, and are
estimated from all serves in the ATP dataset.
We define the serve as the first touch $t=1$, and therefore unforced
errors only occur in rallies corresponding to touches $t=2,3,...$.
From Table 1, we observe a monotonic increasing pattern in the probabilities of the
even number touches $t=2,4,6,8,10$.
This is logical since even number touches correspond to situations where A
was the receiver, and the service advantage to A dissipates as the rally lengthens.
Similarly, we observe a monotonic decreasing pattern in the probabilities of the
odd number touches $t=3,5,7,9$.
This is also logical since odd number touches correspond to situations where A
was the server, and the service disadvantage to A dissipates as the rally lengthens.
We also note that the probabilities converge to the common value 0.573
as the rally lengthens. This also implies that the service advantage eventually
disappears.
It is reasonable that the probabilities 
in Table 1 exceed 0.5 since A has avoided the unforced
error, and this puts B in a position where B must play
the ball and consequently may make a mistake.

Therefore, in the second simulation procedure where an unforced error is made,
we note the touch number where the unforced
error was made, then assume that the
unforced error did not occur, and 
complete the outcome of the rally using the probabilities in Table 1. 

\begin{table}[h]
\begin{center}
\begin{tabular}{r c }
t & \multicolumn{1}{c}{Probability} \\ \hline
2 & 0.535 \\
3 & 0.599 \\
4 & 0.558 \\
5 & 0.586 \\
6 & 0.569 \\
7 & 0.575 \\
8 & 0.571 \\
9 & 0.573 \\
10 & 0.573 \\
\end{tabular}
\caption{Estimated probabilities of player A winning the
point had player A not made an unforced error on touch $t$.
For $t>10$, we use the probabilities corresponding to $t=10$.}
\end{center}
\end{table}

The above algorithm allows us to quantify the impact of
unforced errors in terms of recognizable tennis statistics.
However, we are also able to carry out a what-if
analysis for training purposes. 
Specifically, we can assess the impact of player A reducing their unforced errors
by a given probability, say $x$. 
Here, we adjust the second bootstrapping algorithm
where in a given serve simulation,
with probability $1-x$, we randomly select from the instances 
where player A made an unforced error. In these cases, we
replace the outcome of the serve according to the probabilities
in Table 1.

\section{ANALYSIS}

\subsection{Descriptive Statistics}

As the topic of unforced errors in tennis has not received
much attention in the academic literature, it is instructive
to first explore our dataset using some simple descriptive techniques.
In what follows, we observe some features of unforced errors that have not
previously been investigated.

To begin, we consider all serves in the ATP and WTA datasets
except for first serve faults. Recall that we do not regard
first serve faults as unforced errors since they result in a
second serve opportunity.

Accordingly, consider a rally that involves $t^*$ touches. 
The server would have played the ball $t_s = (t^*+1)/2$ times
if $t^*$ is odd, and 
the server would have played the ball $t_s = t^*/2$ times
if $t^*$ is even. 
Similarly, the receiver would have played the ball $t_r = (t^*-1)/2$ times
if $t^*$ is odd, and 
the receiver would have played the ball $t_r = t^*/2$ times
if $t^*$ is even. 
Therefore, we calculate a player's unforced error rate
as the number of unforced errors made by the player divided
by the number of times that they played the ball. 
This is calculated over all opportunities during a
player's professional career.

In Figure 1, we provide histograms
for the unforced error rate for players in the ATP tour
(Figure 1a) and the WTP tour (Figure 1b).
These histograms consider only players who have played
at least 10 matches.
We observe that 
for most players,
the unforced error
rates in tennis lie in the 
interval (5\%,12\%).
It seems that the unforced error rate is similar for men and women.

\begin{figure}[h]
\centering
\includegraphics[scale = 0.52]{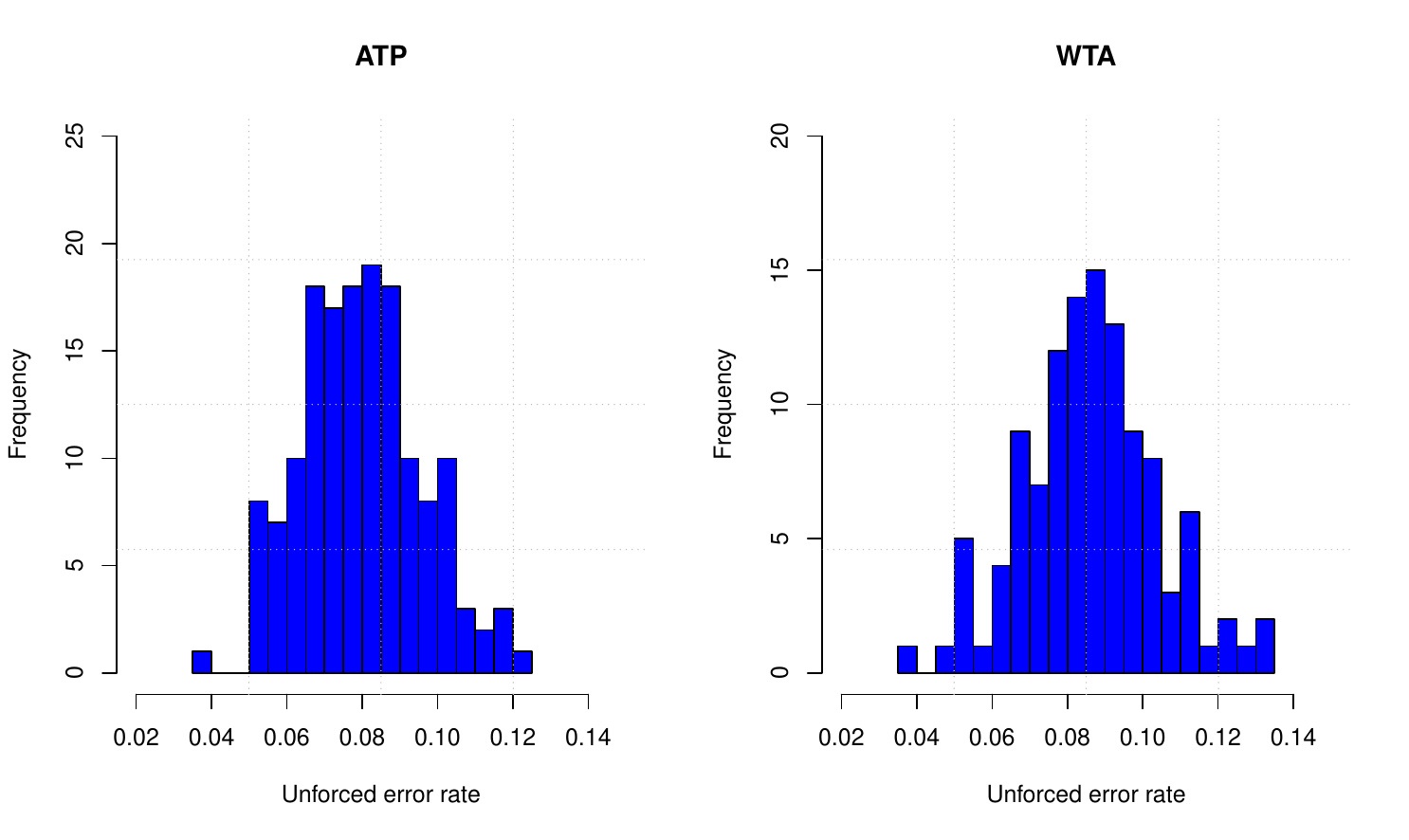}
\caption{Histograms of unforced error rates for players on
the ATP tour and the WTA tour.}
\end{figure}

In Table 2, we provide the five players with the lowest 
unforced error rates and the five players with the highest
unforced error rates. This is done for both the ATP and WTA tours.
On the ATP tour, we observe that the five players with the highest unforced error rates 
have not experienced great success. 
On the other hand, we observe that five players with the lowest unforced error rates 
include some famous players including Mats Wilander and Bjorn Borg.
We also note that Rafael Nadal has the sixth lowest unforced
error rate of 5.4\%.
This provides evidence concerning the importance of minimizing
unforced errors.
Furthermore, amongst the 153 players in the ATP dataset, the 16 players
with the most matches all had unforced error rates less than 8.2\%.
If a player has played a lot of matches, then they have experienced
sustained success. This again emphasizes that low unforced error
rates are related to success.
The same general patterns are observed on the WTA tour where
prominent players Chris Evert and Caroline Wozniacki appear
on the low rate list.

\begin{table}[h]
\begin{center}
\begin{tabular}{c| l c| l c}
Tour & \multicolumn{1}{c}{Player} & Low Rates & \multicolumn{1}{c}{Player} & High Rates \\ \hline 
ATP  & Mats Wilander    & 3.6   & Reilly Opelka        & 12.3 \\
ATP  & Bjorn Borg       & 4.9   & Aslan Karatsev       & 12.1 \\
ATP  & Jaume Munar      & 5.2   & Benoit Paire         & 11.9 \\
ATP  & Gilles Simon     & 5.2   & Peter Gojowczyk      & 11.8 \\
ATP  & Daniel Evans     & 5.3   & Nikoloz Basilashvili & 11.3 \\ \hline
WTA  & Sara Sorribes Tormo  & 4.1   & Jelena Ostapenko     & 13.1 \\
WTA  & Sara Errani          & 4.8   & Madison Keys         & 13.0 \\
WTA  & Monica Niculescu     & 5.0   & Petra Kvitova        & 12.6 \\
WTA  & Chris Evert          & 5.3   & Julia Goerges        & 12.3 \\
WTA  & Caroline Wozniacki   & 5.3   & Dayana Yastremska    & 11.9 \\
\end{tabular}
\caption{Unforced error rate probabilities for players on the ATP
and WTA tours.} 
\end{center}
\end{table}

From Figure 1, we see that on average, the unforced error rate 
across players is roughly
8.0\% (ATP) and 8.6\% (WTA). 
As significant as this statistic may
appear, unforced errors may be viewed as even more
detrimental to success in tennis. 
Recall, every point must terminate with either a forced or an unforced error.
When we look
at the number of serves in tennis that end
by an unforced error, the rates are 
30.3\% and 32.9\%
on the ATP and WTA tours, respectively.

We know that there are different issues in tennis
with respect to the touch number. For example, with big
servers, touch $t=2$ is a great challenge for the receiver
as the main focus is to simply return the serve.
In Figure 2, we plot the unforced error rates for both
the ATP and WTA tours according to touch number.
Given that longer rallies are less common, we have
truncated the plots.
We have also separated the plots according to touches
$t=3,5,\ldots,13$ for the server and touches
$t=2,4,\ldots,12$ for the receiver.
We generally observe that the server's unforced error
rate is steady across the touch number $t$. 
On the other hand, the receiver's
unforced error rate is lower for the initial return $t=2$
and then appears to stabilize.
This may be because the receiver's main objective 
with respect to the serve is to
simply return the serve instead of hitting a winning shot. 
We observe similar patterns
in both the men's and women's tours.

\begin{figure}[H]
	\centering
	\begin{subfigure}[b]{\textwidth} 
		\centering
		\includegraphics[scale=0.4]{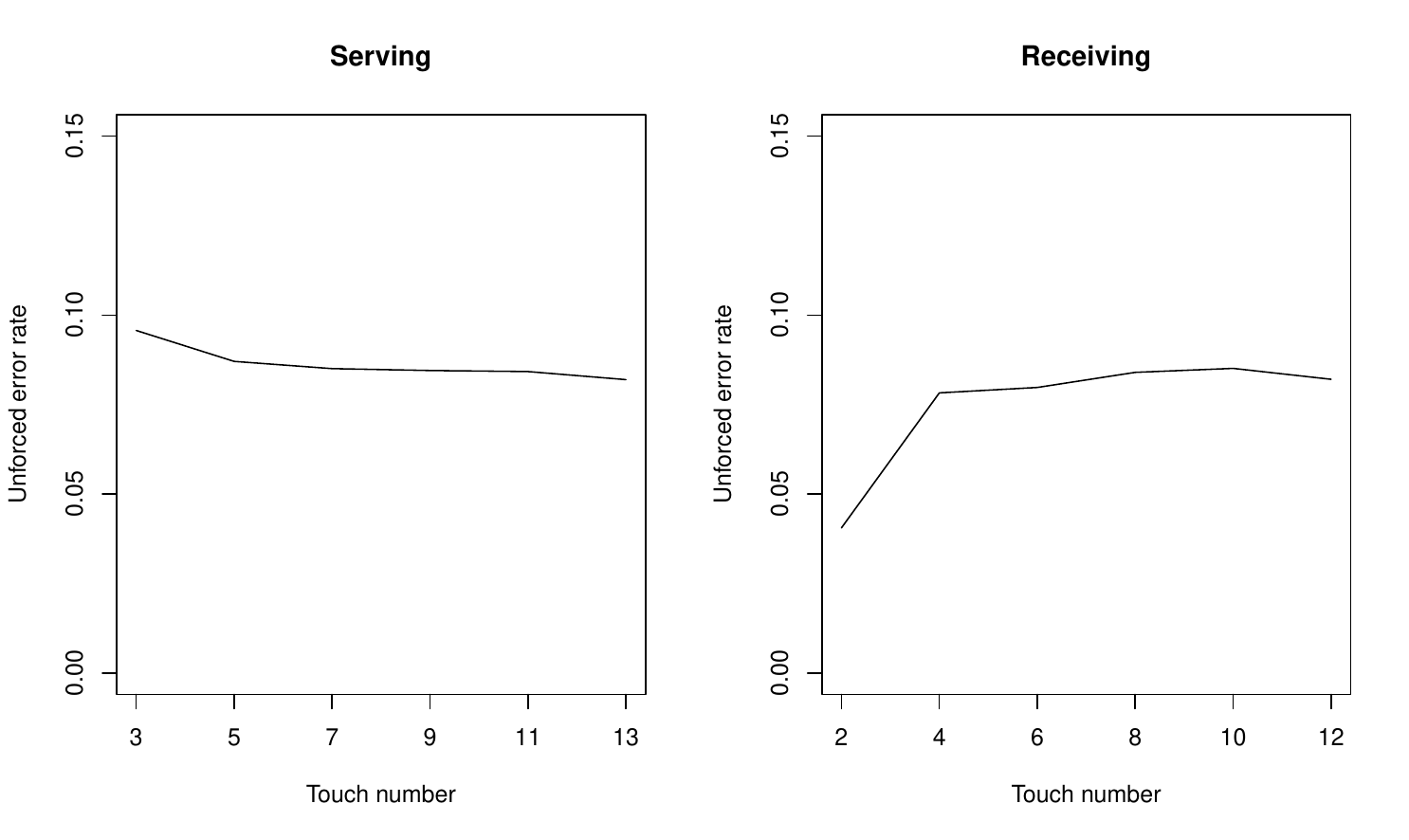}
		\caption{ATP tour}
		\label{fig:atp}
	\end{subfigure}
	\vspace{1em}  
	
	\begin{subfigure}[b]{\textwidth}
		\centering
		\includegraphics[scale=0.4]{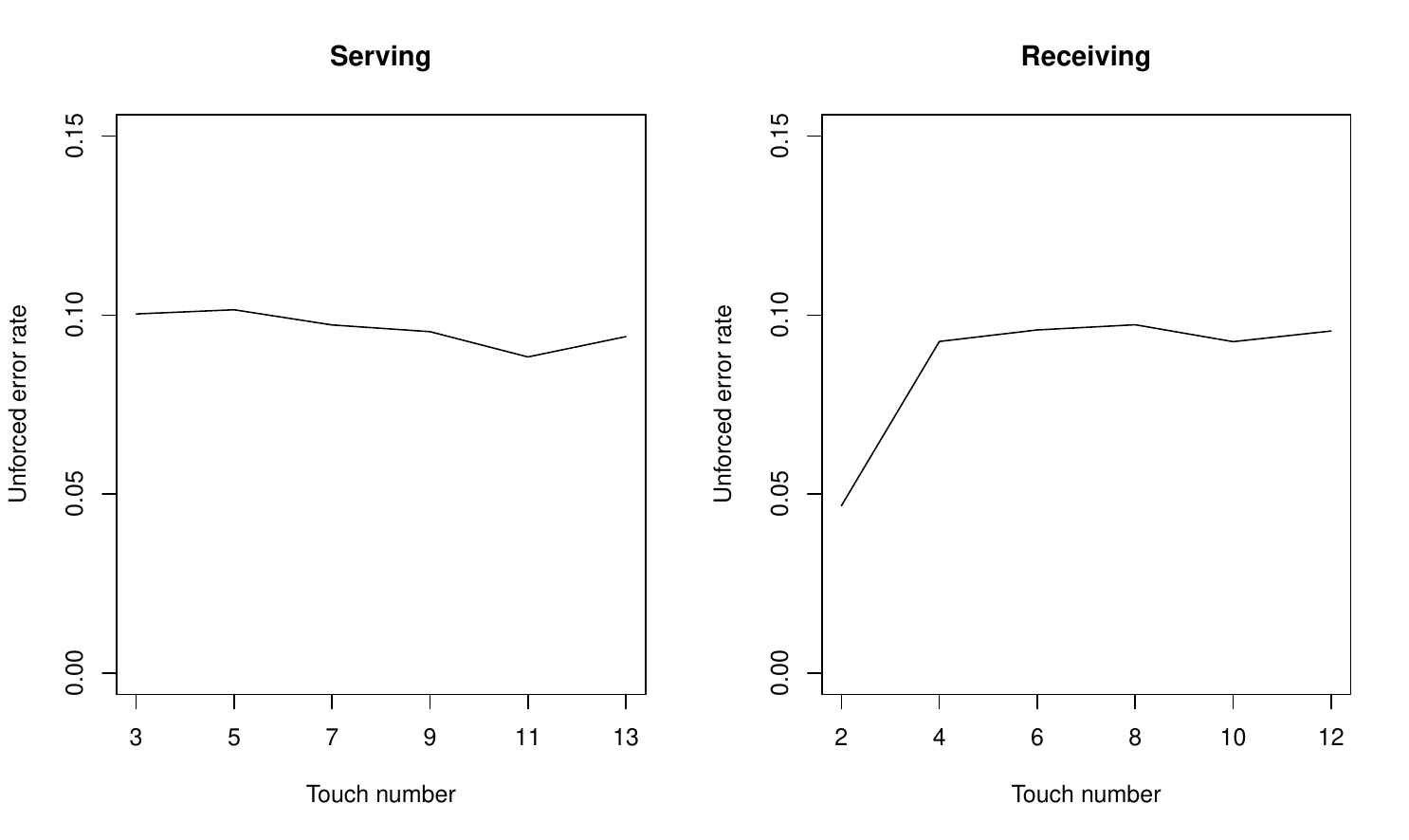}
		\caption{WTA tour}
		\label{fig:wta}
	\end{subfigure}
\caption{Unforced error rates on the ATP and WTA tours with
respect to touch number.}
\label{fig:unforced_errors}
\end{figure}

It is also interesting to investigate how the game has changed
with respect to unforced errors. In Figure 3, we
plot the unforced error rates in the ATP and WTA tours
versus year.
We observe that the unforced error rate has gradually
increased over time in both the men's and women's games.
We do not have a definitive explanation for this.
One conjecture relates to technology changes in tennis,
especially relating to racquets.
With such advances, there has been an increase in the speed
of the game which in turn may make it more difficult to keep
the ball in play, and consequently increase unforced errors.

\begin{figure}[h]
\centering
\includegraphics[scale = 0.5]{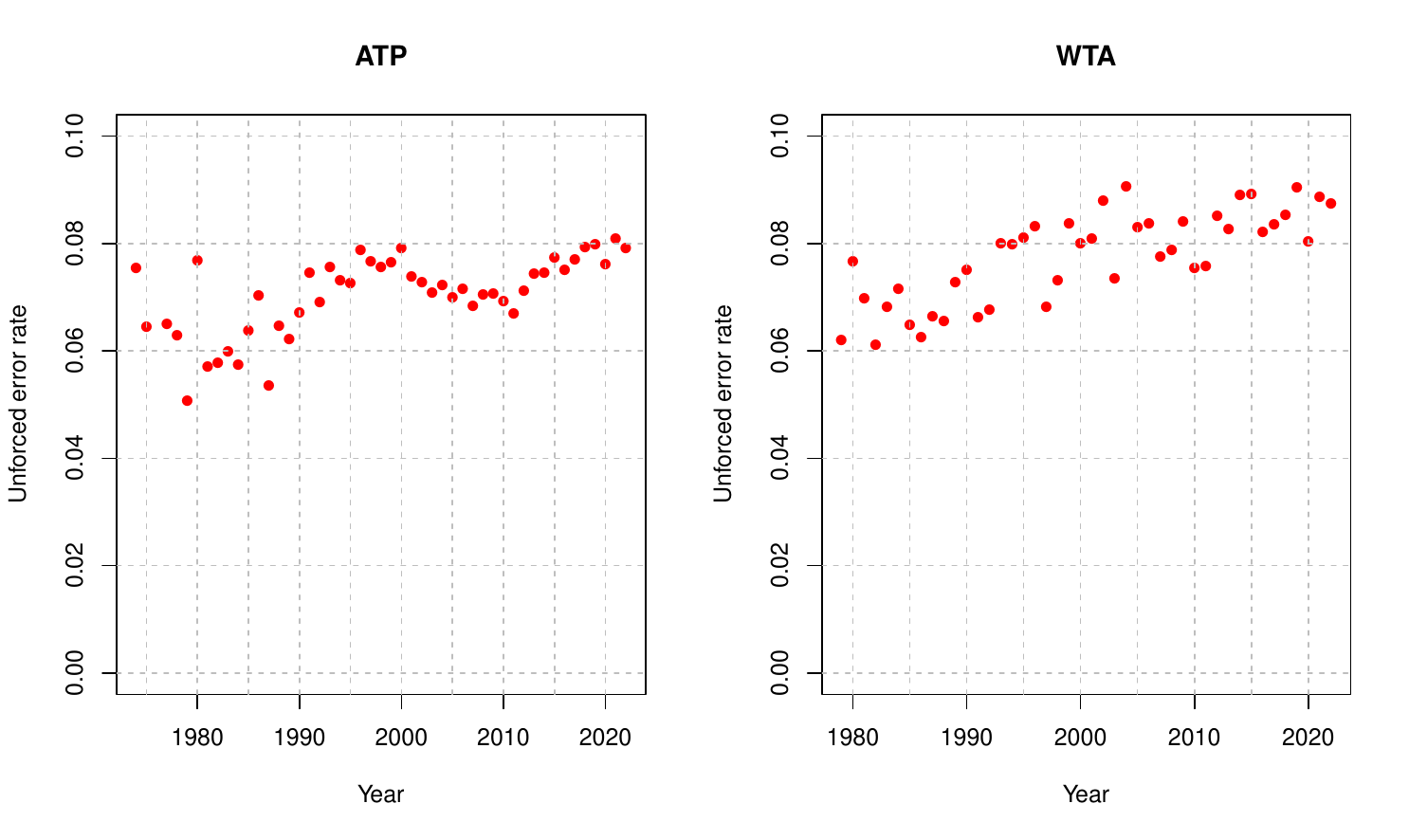}
\caption{Unforced error rates on the ATP and WTA tours
over time.}
\end{figure}

\subsection{Bootstrapping Investigation}

We now illustrate the proposed methods by considering
two prominent players, Roger Federer and Rafael
Nadal. They have enjoyed an intense rivalry, beginning in 2004,
and have played against one another in 9
Grand Slam finals.
In Table 3, we provide some descriptive comparisons of the two players.
In comparison with Figure 1, we note that Federer's career unforced error
rate of 8.2\% is about average for an ATP player whereas
Nadal's rate of 5.4\% is extremely low.
We note that Nadal has made tentative plans to retire
in 2025.

\begin{table}[h]
\begin{center}
\begin{tabular}{l| c c c c}
Player & ATP Start (Age) & ATP End & ATP Matches & UFE \\ \hline
Roger Federer & 1998 (17) & 2022        & 1526  & 0.082 \\
Rafael Nadal  & 2001 (14) & 2024-active & 1300+ & 0.054 \\
\end{tabular}
\caption{Information concerning the singles careers of Federer and Nadal
with career unforced error rates (UFE).}
\end{center}
\end{table}

We now carry out an analysis involving Federer and Nadal with the results
shown in Table 4.
We show actual results between the two players, simulated
results using Federer's historic
unforced error rate, 
simulated results with Federer's unforced error
rate reduced by 10\% and
simulated results with Federer's unforced errors eliminated
(i.e.\ reduced by 100\%).
These results are based on the 35 ATP matches recorded between
Federer and Nadal in our dataset.
We note that official records indicate that these two players actually faced
each other 40 times; hence the dataset is missing 5 matches.
In the simulations, we do not use the career unforced error rates.
Instead, we use the unforced error rates (UFE) exhibited between the two
players in the 35 matches. In these matches,
Federer's UFE was 8.9\% and
Nadal's UFE was 5.7\%. 
The inter-personal UFE rates between
the two players are very close to their career rates.
Also, for simulated match results, a win is defined as the
first player to have won three sets.
The results in Table 4 are based on simulations from $N= 3000$
matches that took roughly two hours on a laptop computer.

From Table 4, in actual competitions,
we observe that Federer won 49.3\% of
the points against Nadal, demonstrating the competitiveness
of the matches. However, at the match level, Federer won
only 16 of the 40 matches, i.e.\ only 40\% of the time. 
This demonstrates something about Nadal's ability to come
through in clutch situations.
This is further confirmed by Federer's simulation match
results which are better than his actual match results.
We have confidence in the simulation
results since the differences between Federer's actual
points won and simulated points won is minor. 
Most importantly, we quantify the extent to which Federer's
unforced errors impacted his results against Nadal.
Had Federer been able to reduce his unforced error rate
by even a small margin (10\%), his match results would
have increased substantially to 53.7\%, placing him above Nadal.
And as an upper bound for improvement, we consider
the simulation case where Federer eliminated all unforced
errors. In this unrealistic scenario, Federer would have been
completely dominant. This exercise illustrates the critical
impact of unforced errors in tennis.

\begin{table}[h]
\begin{center}
\begin{tabular}{l| l l l l}
\multicolumn{1}{c|}{Cases} & Points Won & Games Won & Matches Won \\ \hline
Federer - Actual              & 49.3 & 47.7 & 40.0 \\
Federer - Simulated Historic UFE Rate   & 49.5 (0.07) & 48.9 (0.18) & 44.5 (0.90) \\
Federer - Simulated 10\% UFE Reduction  & 50.3 (0.07) & 50.7 (0.18) & 53.7 (0.91) \\
Federer - Simulated 100\% UFE Reduction & 58.6 (0.08) & 67.3 (0.20) & 98.7 (0.21) \\
\end{tabular}
\caption{Federer percentage results against Nadal under various scenarios.
Bootstrap standard errors
are given in parentheses.}
\end{center}
\end{table}

\newpage

\section{DISCUSSION}

Whereas unforced errors in tennis are a much discussed topic,
the quantification of the impact of unforced errors is something
that appears to be missing from the literature. This paper uses
bootstrapping methods to address this novel problem. The consequent 
understanding of a player's unforced error rate can then be used
for player evaluation and training.

In Section 4.2, we investigated the the impact of unforced errors
between two giants of men's tennis, Roger Federer and Rafael Nadal.
This was a convenient investigation since these two players had an
extensive history of playing against one another, and hence,
their unforced error rates could be well-assessed. 
In future investigations, one may wish to study pairs of players with
less of a competitive history. In these cases, one could use their
performances against the field (i.e.\ the typical player) to assess
unforced error rates. Doing so would allow us to compare any
pair of players, even those from different eras.

Another avenue for future investigation may involve digging
deeper in the circumstances in which unforced errors are
committed. In live matches, unforced error rates are routinely
reported, but knowing when these errors occur is relevant
to performance.
For example, they may occur on high or low leverage points,
early or late in a rally, etc. 

\newpage

\section{REFERENCES}
	
\begin{description}
\begin{small}
		
\item Albert, J.A., Glickman, M.E., Swartz, T.B.\ and Koning, R.H., Editors (2017).
{\em Handbook of Statistical Methods and Analyses in Sports},
Chapman \& Hall/CRC Handbooks of Modern Statistical Methods,
Boca Raton.

\item Attray, H.\ and Attray, S.\ (2021).
Cost of an unforced error in tennis - A statistical approach.
{\em International Tennis Federation}, Issue 84, 12-14.

\item Baker, R.D.\ and McHale, I.\ (2014).  A dynamic paired comparisons model: 
Who is the greatest tennis player?
{\em European Journal of Operational Research}, 236 (2), 677-684.

\item Baker, R.D.\ and McHale, I.\ (2017).
An empirical Bayes model for time-varying paired comparisons rankings:
Who is the greatist women's tennis player?
{\em European Journal of Operational Research}, 258 (1), 328-333.

\item Barreira, J.\ and Chiminazzo, J.G.C.\ (2020).
Who, how and when to perform winner points and unforced errors in
badminton matches? An analysis of men's single matches in the 2016
Olympic Games.
{\em International Journal of Performance Analysis in Sport}, 20(4), 610-619.

\item Brody, H.\ (2006).
Unforced errors and error reduction in tennis.
{\em British Journal of Sports Medicine}, 40(5), 397-400.


\item Epasinghege Dona, N., Gill, P.S.\ and Swartz, T.B.\ (2024). 
What does rally length tell us about player characteristics in tennis?
{\em Journal of the Royal Statistical Society, Series A},
https://doi.org/10.1093/jrssa/qnae027

\item Gudmundsson, J.\ and Horton, M.\ (2017).
Spatio-temporal analysis of team sports.
{\em ACM Computing Surveys}, 50(2), Article 22.

\item Klassen, F.J.G.M.\ and Magnus, J.R.\ (2001).
Are points in tennis independent and identically distributed? Evidence
from a dynamic binary panel model.
{\em Journal of the American Statistical Association}, 96(454), 500-509.

\item Klassen, F.\ and Magnus, J.R.\ (2014).
{\em Analyzing Wimbledon: The Power of Statistics},
University Press Scholarship Online, 
https://doi.org/10.1093/acprof:oso/9780199355952.001.0001

\item Kovalchik, S.A.\ and Albert, J.\ (2022).
A statistical model of serve return impact patterns in professional tennis.
Accessed October 13, 2022 at https://arxiv.org/abs/2202.00583

\item McHale, I.\ and Morton, A.\ (2011).
A Bradley-Terry type model for forecasting tennis match results.
{\em International Journal of Forecasting}, 27 (2), 619-630.

\item Mecheri, S., Rioult, F., Mantel, B., Kauffmann, F.\ and Benguigui, N.\ (2016).
The serve impact in tennis: First large-scale study of big hawk-eye data.
{\em Statistical Analysis and Data Mining}, 9(5), 310-325.

\item Tea, P.\ and Swartz, T.B. (2022).
The analysis of serve decisions in tennis using Bayesian hierarchical models.
{\em Annals of Operations Research}, 325(1), 633-648.

\item Wei, X., Lucey, P., Morgan, S., Carr, P., Reid, M.\ and Sridharan, S.\ (2015).
Predicting serves in tennis using style priors.
{\em Proceedings of the 21th ACM SIGKDD International Conference on Discovery and Data Mining}, 2207-2215.

\item Yadav, S.K.\ and Shukla, Y.M.\ (2011). 
Analysis of unforced errors in relation to performance in singles in badminton.
{\em International Journal of Physical Education}, 4(2), 117-119.
 
\end{small}
\end{description}

\end{document}